# Dual-Band, Slant-Polarized MIMO Antenna Set for Vehicular Communication

Rasool Keshavarz[1], Dan Winson[2], Justin Lipman[1], Mehran Abolhasan[1], and Negin Shariati[1,3]

[1] RF and Communication Technologies (RFCT) Research Laboratory, University of Technology Sydney, Australia
rasool.keshavarz@uts.edu.au, negin.shariati@uts.edu.au
[2] Zetifi, 1/1 Riedell St, Wagga Wagga, NSW 2650, Australia, dan@zetifi.com
[3] Food Agility CRC Ltd, 175 Pitt St, Sydney, NSW 2000, Australia

*Abstract*— Slant-polarized Multi Input Multi Output (MIMO) antennas are able to improve the performance of mobile communication systems in terms of channel capacity. Especially, the implementation of MIMO configurations for automotive applications requires to consider high gain, wideband, low-profile and affordable antennas in the communication link. In this work design, simulation and measurement of a new dual-band slant-polarized MIMO antenna with HPBW (Half Power Beam Width) of around $90^0$ are presented. Then, four replicas of the proposed antenna set are placed at four different poles (North, South, West and East) to cover $360^0$ around the vehicle as an omni-directional pattern. In the real world scenario, the proper antenna set is selected to communicate with the intended user. Each slant MIMO antenna set consists of two inclined ($45^0$) low band (LB: 700 to 900 MHz) and two inclined high band (HB: 1.7 to 2.7 GHz) log-periodic antennas. The measured gain of LB and HB antennas are 7 dBi and 8 dBi, respectively. Great agreement between simulation and measurement results confirms the accuracy of the design and simulation procedures of antenna system using optimization algorithm (Genetic method). The proposed antenna is also measured in the field for industrial applications.

*Index Terms*— Connectivity, Log-periodic antenna, measurements, MIMO, slant polarization, vehicle antenna.

## I. INTRODUCTION

Within the framework of vehicular communication systems, the requirement of high data rate, reliable, low power and low-profile links are inevitable [1]. An admirable solution for robust fast link is MIMO scenario, taking advantage of spatial diversity with multiple antennas at both transmitter and receiver [2]. The benefits yielded from spatial diversity increase with the number of antennas employed. However, there is the tradeoff between transmission rate and spatial diversity. The multitude of antennas elements can either send different data over different paths (multiplexing gain), or the same signal over different paths (diversity gain) [3]. Commonly known as the diversity-multiplexing tradeoff, there is a balancing act between these two methods of transmission.

In MIMO paradigm, wideband and compact antennas perform a vital role in mobile communications which has been attracting significant interests from researchers [4], [5]. New wideband, high gain compact and highly isolated antennas are essential in the MIMO system [6]. Moreover, various studies show the ability of slant polarization to provide signal improvement through fading communication channel as well as in NLOS conditions for urban and other environments compared with horizontal and vertical polarizations. Therefore, slant MIMO antennas are a good candidate for reliable and high-speed communication links [7].

Existing vehicular communications systems lack the capabilities required to deliver the fast and reliable connectivity needed for Connected and Autonomous Vehicle (CAV) applications. This issue is particularly prevalent in rural and regional locations where the signal from distant cellular network towers is inadequate when used with legacy antennas that are typically omni-directional SISO antennas or incorrectly polarized MIMO arrays [8].

Benchmarks for selecting the best antenna for vehicular communication scenario are gain (increasing range and SNR), dimension (flexible in usage), fabrication (easy manufacturing), cost, polarization (resistance against fading), isolation, azimuth angle (omni coverage) and elevation angle (coverage in height). Full coverage ($360^0$) performance of the vehicle antenna with an acceptable gain (>6 dBi) within operating frequency (LTE and WiFi) leads to a large antenna which is not practical for vehicular communication system [8], [9]. Hence, using a number of directive antennas with switching capability between them is a good candidate to realize an omni-directional antenna system.

In this paper, based on the theoretical analysis, simulation results and industry requirements, dual-band log-periodic MIMO antennas are designed and four of them are placed in slant polarization to constitute an omni-directional pattern to cover $360^0$ (four poles; North, South, West and East) around the vehicle. Section II of this paper describes the proposed Dual-band MIMO antenna design, Section III illustrates the simulated and measured results of the proposed antennas and Section IV is conclusion.

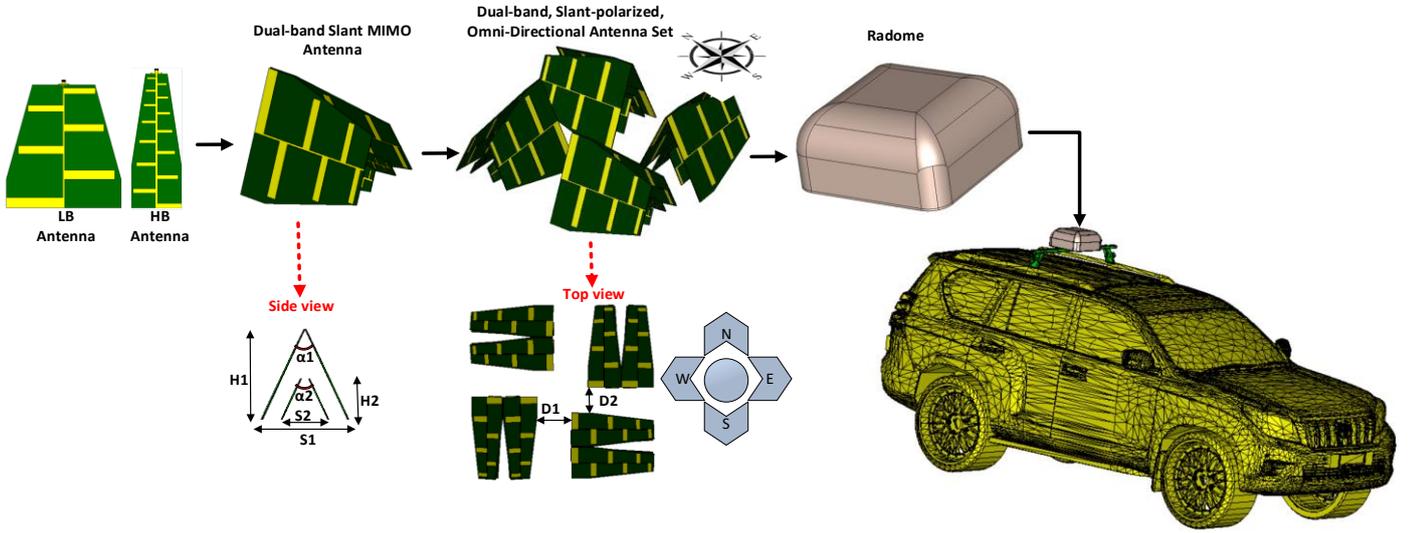

Fig. 1. Design procedure of dual-band slant-polarized antenna set in MIMO configuration for omni-directional vehicular communications.

## II. DUAL-BAND SLANT-POLARIZED MIMO ANTENNA

Fig. 1 shows the proposed approach for designing dual-band slant-polarized antenna set in MIMO configuration to achieve omni-directional pattern. According to this figure, first step is designing low band (LB) and high band (HB) antennas to cover 700 MHz to 900 MHz and 1.7 GHz to 2.7 GHz (LTE and WiFi), respectively. Then, the antennas are considered as slant-polarized MIMO configuration to reach omni-directional pattern (four antennas to cover all poles). Finnaly, a radome is designed for the whole antenna set.

### A. LB and HB Log-Periodic Antenna (LPA) Design

The proposed antenna element is log-periodic antenna (LPA) which consists of a transmission line to drive a number of dipoles. Each dipole covers a specific frequency band and the whole structure constitutes a wide-band antenna which the spacing between dipoles, length and width of dipoles determine the performance of the LPA (gain, bandwidth, etc.) [8], [9]. Fig. 2 exhibits the schematic of a LPA where $L_i$ and $W_i$ show length and width of $i^{th}$ dipole, respectively and $d_i$ is the distance between two adjacent dipoles ($i=1, 2, ..., n$). In this antenna, a linearly polarized beam is observed in the direction of smaller elements. For any frequency within the design band, there are some elements, which are nearly half wavelength. The relationship between the parameters shown in Fig. 2 can be summarized as follows:

$$\alpha = \frac{d_i}{d_{i-1}} = \frac{L_i}{L_{i-1}} \quad (i=1, 2, ..., n) \quad (1)$$

where $0.78 < \alpha < 0.98$ is the geometric ratio [8].

Since the total length of LPA is very large to cover 700 MHz to 2.7 GHz and it is not practical for vehicle communication system, in this paper, two separate LB (700 MHz to 900 MHz) and HB (1.7 GHz to 2.7 GHz) antennas are designed and placed as slant-polarized configuration in a MIMO antenna set. The design procedure of MIMO antennas and omni-directional antenna set are presented in Section B.

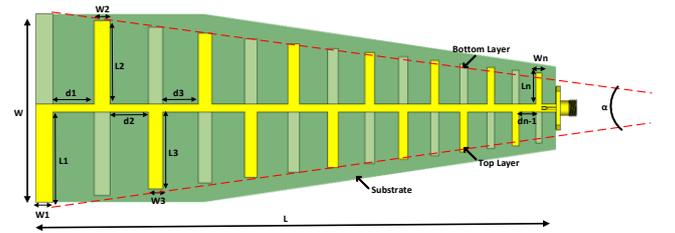

Fig. 2. Layout of log-periodic antenna (LPA) where n is the total number of poles.

### B. Omni-directional Antenna Set

Due to the mutual coupling between LB-LB, HB-HB and LB-HB antennas in the slant-polarized MIMO antenna, the design process of the omni-directional antenna set in Fig. 1 is not straightforward and needs optimization methods. According to Fig. 3, the input parameters of the optimization problem are gain, HPBW, axial ratio, operating frequency and dimension. The optimization algorithm, (Genetic method in CST software) extracts the optimum LB and HB antennas based on the input parameters which are used in the simulation process of omni-directional antenna set. The simulation results of the whole antenna set are considered as a feedback line in the design process and make amendment in the LB and HB antenna design ($L_i$, $W_i$ and $d_i$). The design loop is terminated when the desired goals are encountered with an acceptable error value. The total error of design procedure, $e_t$, is calculated as:

$$e_t = w_f e_f + w_g e_g + w_p e_p + w_d e_d + w_h e_h \quad (2)$$

Where $e_f$, $e_g$, $e_p$, $e_d$ and $e_h$ are errors of operating frequency, antenna gain, axial ratio (polarization), dimension and HPBW and $w_f$, $w_g$, $w_p$, $w_d$, and $w_h$ are their associated weighting functions in the error calculation.

In MIMO structure of dual-band slant-polarized antennas, the maximum gain occurs at $\alpha=45^0$ (pure slant polarization) when polarization loss factor (PLF) is minimum. α is the angle between two antennas in the slant MIMO set in Fig. 1.

$$PLF = \cos(\alpha/2 - 45^0) \quad (3)$$

On the other hand, there is a trade-off between the height of antenna set and PLF. According to Fig. 4, by increasing α from $45^0$, the height of MIMO set reduces while PLF increases, leading to signal loss in the Tx or Rx mode.

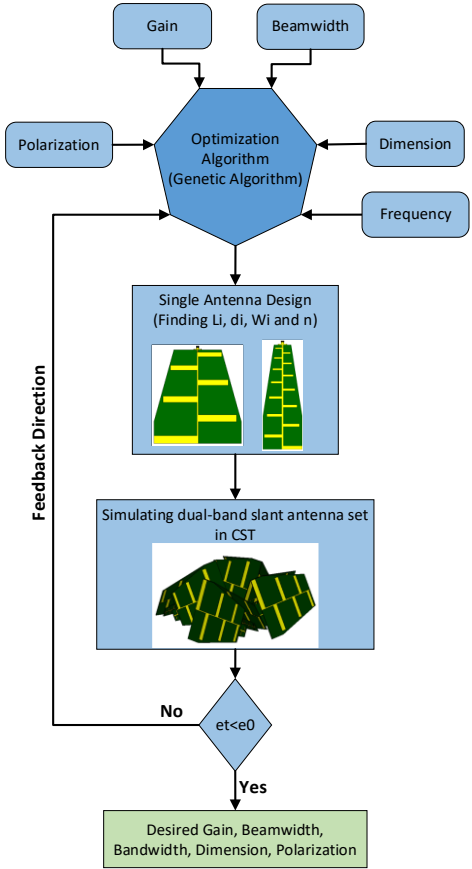

Fig. 3. Design procedure of the proposed dual-band slant-polarized antenna in MIMO configuration using optimization algorithm in CST.

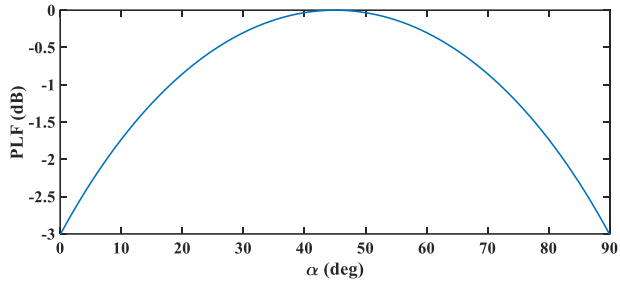

Fig. 4. Polarization loss factor (PLF) of slant-polarized MIMO antenna versus angle between two antennas (α in Fig. 1).

## III. SIMULATION AND MEASUREMENTS

The proposed LB and HB antennas are fabricated on low cost FR-4 substrate with permittivity of 4.3, thickness of 1.6 mm and $\tan(\delta) = 0.03$ (Fig. 4). Dimension of the fabricated prototypes is presented in Table I.

Fig. 5 shows the simulated and measured return loss of the fabricated antennas and the testing setup for this measurement. According to Fig. 5, simulation and measurement results are in great agreement and the $|S_{11}|$ is better than -10 dB and -15 dB within the operating frequency range of LB and HB antennas, respectively.

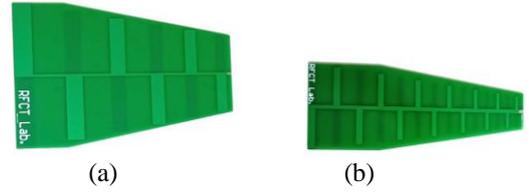

Fig. 4. Fabricated prototypes of the proposed antennas on FR-4 substrate ($\varepsilon_r = 4.3, h = 1.6\ mm, \tan(\delta) = 0.03$), a) low band (LB), b) high band (HB).

TABLE I. DIMENSION OF LB AND HB ANTENNAS

| Ant | Li (mm) | Wi (mm) | Di (mm) | L (mm) | W (mm) |
|---|---|---|---|---|---|
| LB | 87, 77, 68, 60, 53, 47 | 16, 14, 12, 11, 10, 9 | 28, 25, 22, 20, 17 | 190 | 176 |
| HB | 33, 30, 28, 26, 24, 22, 20, 19, 17, 16, 15, 13, 12, 11 | 12, 11, 11, 10, 10, 10, 10, 9, 9, 9, 8, 8, 7, 7 | 15, 14, 14, 13, 13, 13, 12, 12, 12, 11, 11, 10, 10, 9 | 190 | 69 |

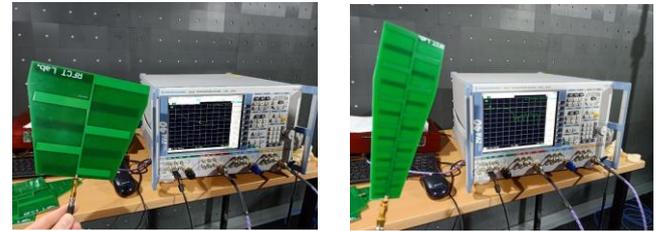

(a)

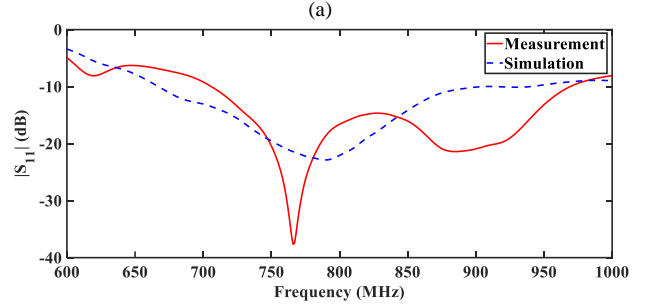

(b)

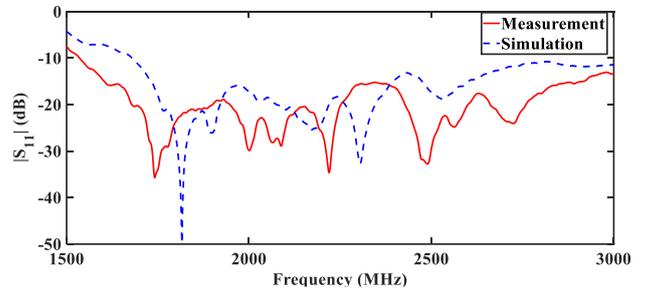

(c)

Fig. 5. a) Measurement setup, and simulated and measured rreturn loss of the proposed antennas, b) LB and, c) HB.

Then, the slant-polarized MIMO configuration is implemented using two HB and two LB antennas as Fig. 1.

Since the limiting factor on the height of antenna system is LB antenna, we set $\alpha_2 = 45^0$ to reach PLF=0 dB for HB MIMO antennas. Therefore, according to Table I, the height of HB MIMO antennas ($H_2$) equals:

$$H_2 = W_{HB} \times \cos(\alpha_2) = 69 \times \cos(45) = 48.7\ mm \quad (4)$$

and the maximum distance between antennas in HB MIMO set is given as:

$$S_2 = 2 \times W_{HB} \times \sin(\alpha_2) = 97.5\ mm \quad (5)$$

Therefore, the height ($H_1$) and maximum distance ($S_1$) of LB MIMO antennas should be more than 48.7 mm and 97.5 mm, respectively. In Fig. 6, the width and height of the LB MIMO antennas are calculated for different $\alpha_1$ values and due to $S_2 = 97.5$ and $H_2 = 48.7$, the red zone in this figure cannot be considered in the design process. Moreover, Fig. 4 and Fig. 6 illustrate the trade off between MIMO antenna set height and PLF value.

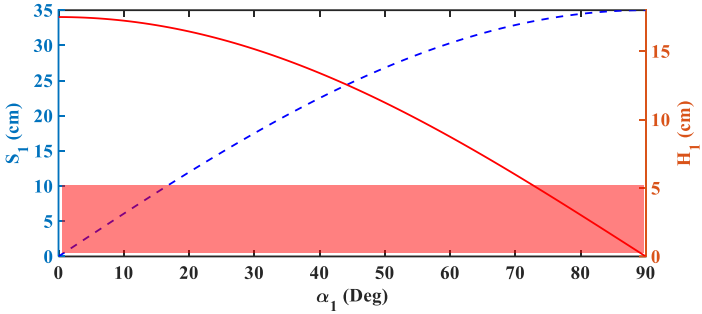

Fig. 6. Dimension of MIMO slant antenna set versus incline angle ($\alpha_1$).

The simulated and measured radiation pattern (normalized realized gain) of the single antennas (LB and HB) are presented in Fig. 7. This figure illustrates that the HPBW of LB antenna at 800 MHz on the H-plane and E-plane are $97^0$ and $64^0$, respectively. Moreover, HB antenna shows HPBW of $85^0$ and $61^0$ at 2.4 GHz for H-plane and E-plane, respectively. The HPBW of the proposed LB and HB antennas on H-plane are acceptable to cover $360^0$ around the final antenna set which is fabricated using 4 dual-band slant-polarized MIMO antennas at four poles (N, S, W, and E).

Fig. 8 exhibits fabricated slant-polarized MIMO antennas, the structure of omni-directional antenna set and the fabricated ABS radome using 3D printer. In the design process of radome the aerodynamics, minimum effect on the radiation pattern, low cost, light weight, for vehicles and durability are considered. Furthermore, a wooden mounting structure is designed and fabricated to install the antenna set under test in the measurement setup (Fig. 8(e) and (f)). Fig. 9 and Fig. 10 show the simulation and measurement results of slant-polarized MIMO antenna and omni-directional antenna set, respectively. Maximum realized gain of LB and HB antennas in the omni-directional antenna set are presented in Fig. 11 and the maximum gain of LB and HB antennas are 7.8 dBi and 8.2 dBi, respectively.

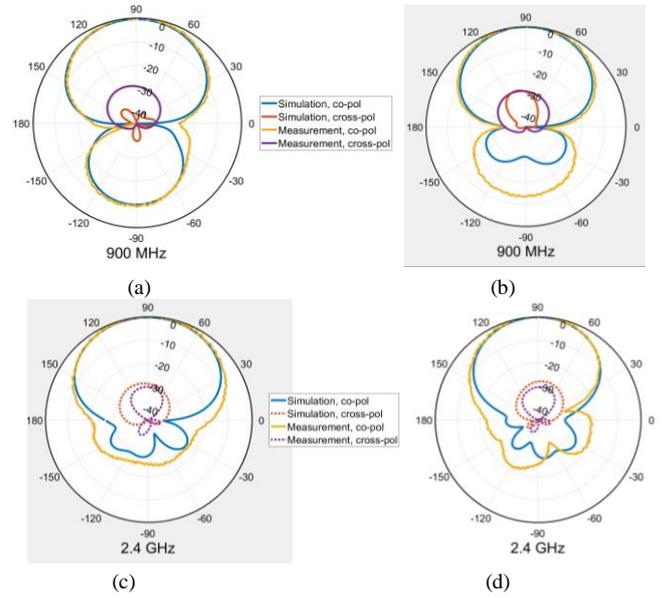

Fig. 7. The radiation pattern (normalized value of realized gain) of single antennas, a) LB, E-plane, b) LB, H-plane, c) HB, E-plane, d) HB, H-plane.

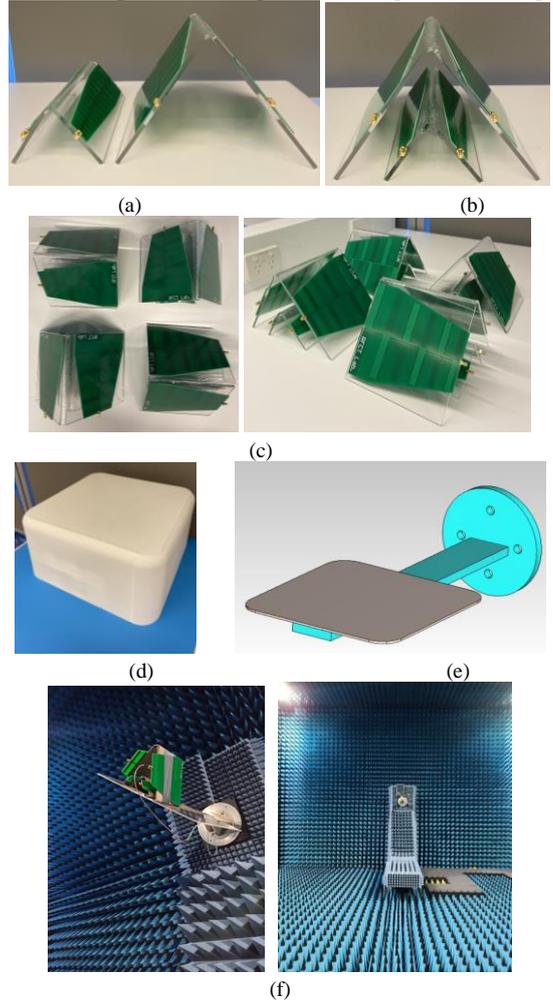

Fig. 8. Fabricated prototypes and measurement setup, a) LB and HB antennas, b) dual-band slant-polarized MIMO antennas, c) omni-directional antenna (top and perspective views), d) radome, e) wooden mounting structure, f) measurement setup.

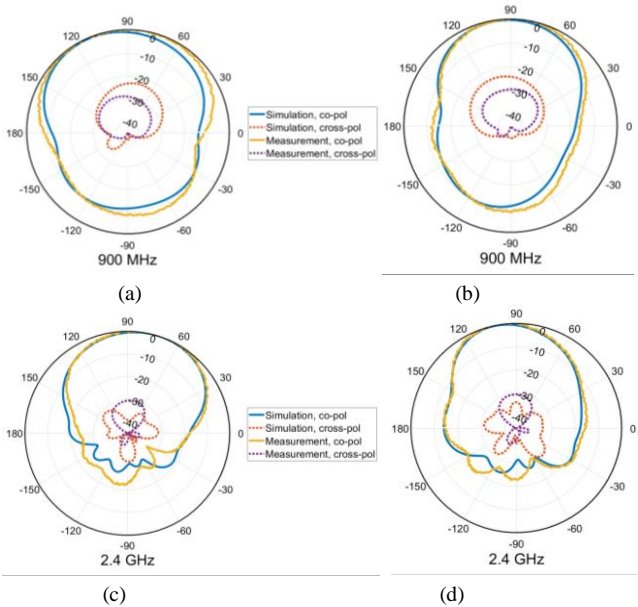

Fig. 9. The radiation pattern (normalized value of realized gain) of dual-band MIMO slant-polarized antennas, a) LB, E-plane, b) LB, H-plane, c) HB, E-plane, d) HB, H-plane.

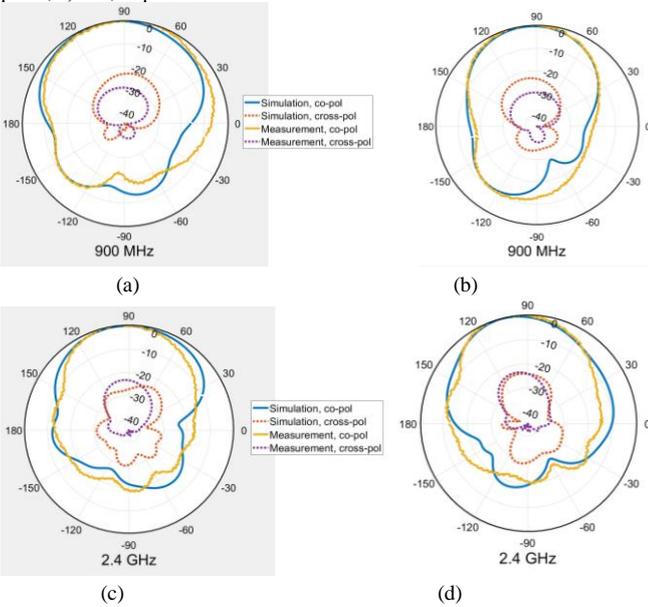

Fig. 10. The radiation pattern (normalized value of realized gain) of omni-directional antenna set, a) LB, E-plane, b) LB, H-plane, c) HB, E-plane, d) HB, H-plane.

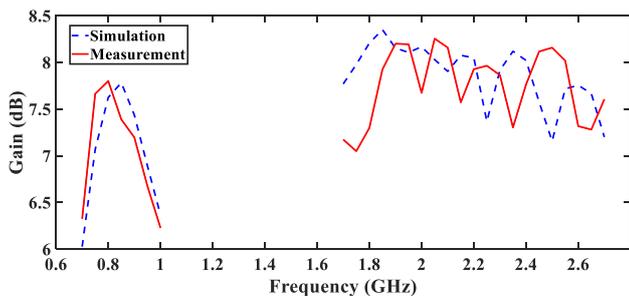

Fig. 11. Maximum gain of LB and HB antennas versus frequency in omni-directional antenna set.

## IV. Conclusion

This work presented the design procedure of dual-band omni-directional slant-polarized MIMO antenna set for vehicular communication system. The design method is based on optimization algorithm in CST (Genetic method). The antennas are fabricated and different configurations of MIMO set with different incline angles ($\alpha$) are measured. Then, we compared them in terms of gain, bandwidth, return loss, and size and found the optimum solution for predefined communication scenario. The main advantages of the proposed MIMO antenna set are compact size, low cost, good isolation between antennas, capability to embed the switching board and GPS antenna in the radome with minimal effect on the antenna performance and simple installation on vehicles.

## Acknowledgment

In undertaking this work, we would like to acknowledge the support of the AusIndustry Entrepreneurs' Programme and Zetifi Company. This project was funded by the Department of Industry, Science and Resources, Innovation Connections grant and Zetifi (Agsensio Pty Ltd), an Australian wireless networking startup. We would also like to acknowledge the support of Food Agility Cooperative Research Centre (CRC) Ltd, funded under Commonwealth Government CRC Program to continue this project. The CRC Program supports industry-led collaboration between industry, research and the community.